\documentclass[aps,prb,twocolumn,superscriptaddress,showpacs,ams]{revtex4}

\usepackage{graphicx}
\usepackage{bm}
\usepackage{amssymb} 
\usepackage{psfrag,color} 

\newcommand{\beq}   {\begin{equation}}
\newcommand{\eeq}   {\end{equation}}
\newcommand{\ba}   {\begin{eqnarray}}
\newcommand{\ea}   {\end{eqnarray}}

\newcommand{\dotleads}   {\mbox{\scriptsize Dot-Leads}}
\newcommand{\leads}   {\mbox{\scriptsize Leads}}

\newcommand{\imp} {\mbox{\scriptsize imp}}

\begin{document}

\title{Phonon-assisted tunneling and two-channel Kondo physics in molecular junctions. }
\author{Luis~G.~G.~V. Dias da Silva}
\affiliation{Materials Science and Technology Division,
Oak Ridge National Laboratory, Oak Ridge, Tennessee 37831, USA and\\
Department of Physics and Astronomy, University of Tennessee, Knoxville, Tennessee 37996, USA}

\author{Elbio Dagotto}
\affiliation{Materials Science and Technology Division,
Oak Ridge National Laboratory, Oak Ridge, Tennessee 37831, USA and\\
Department of Physics and Astronomy, University of Tennessee, Knoxville, Tennessee 37996, USA}
\date{\today}

\begin{abstract}

The interplay between vibrational modes and Kondo physics is a
fundamental aspect of transport properties of correlated molecular
conductors. We present theoretical results for a single molecule
in the Kondo regime connected to left and right metallic leads,
creating the usual coupling to a conduction channel with
left-right parity (``even"). A center-of-mass vibrational mode
introduces an additional, phonon-assisted, tunneling through the
antisymmetric (``odd") channel. A non-Fermi liquid fixed point,
reminiscent of the two-channel Kondo effect, appears at a critical
value of the phonon-mediated coupling strength. Our numerical
renormalization-group calculations for this system reveal
non-Fermi-liquid behavior at low temperatures over lines of
critical points. Signatures of this strongly correlated state are
prominent in the thermodynamic properties and in the linear
conductance.

\end{abstract}

\pacs{71.10.Hf, 72.15.Qm,73.23.Hk,73.63.-b}

%
%
%

 \maketitle
\section{Introduction}
\label{sec:Intro}

Ground-breaking experimental results in single molecular
transistors during the last decade\cite{Natelson::2006a} have
greatly expanded the field of molecular electronics, opening
several possibilities for technological applications and
investigations of fundamental aspects of the physics of these
devices. By now it is established that strong correlation effects
play a key role in the electronic transport through these systems,
as evidenced by the observation of the Kondo
effect\cite{HewsonBook} in both break junctions
\cite{KondoBreakJunctions,Yu:256803:2005} and STM setups.
\cite{KondoSTM}
%
A clear understanding of the mechanisms involved in the emergence
of the Kondo effect in molecular systems is thus of primary
importance.

One possibility to advance our knowledge of these
devices
is by establishing analogies with the well-known transport
properties of semiconductor quantum dots in the Kondo regime.
This, however, proves to be a challenging approach for several
reasons: molecule-leads couplings are very sensitive to the
particular configurations, charging energies are significantly
larger and, more importantly, deformations and vibrational modes
in the molecule play an active role in transport.
\cite{BreakJunctionsPhonons}

This variety of competing effects also brings theoretical
challenges, such as the interesting prospect of investigating the
interplay between vibronic states and
Kondo physics.
Different studies have investigated the effect of
electron-phonon couplings in the charge degrees of freedom of the
molecule, affecting the exchange correlations leading to the Kondo
effect. \cite{NRGphonons,Paaske:176801:2005,Cornaglia:241403:2007}
This issue has also been highlighted in recent experiments
reporting anomalous behavior in the Kondo
transport\cite{Yu:256803:2005} which have been attributed to the
``dressing" of the local energies by Holstein-like phonons.
\cite{Cornaglia:241403:2007}

In addition to these local effects, considerable attention has
been given to effects of vibrational modes in the
\textit{tunneling} from the molecule to the leads. Phonon-assisted
couplings by
``breathing''\cite{Al-Hassanieh:256807:2005,Cornaglia:075320:2005}
or
``center-of-mass''\cite{Al-Hassanieh:256807:2005,BalseiroCM,Mravlje:205320:2006,Mravlje::2008}
molecular modes create  additional correlations with the electrons
in  the
%
%
leads. Such phonon-mediated tunneling processes will, in general,
lead to novel features in the transport properties
\cite{Al-Hassanieh:256807:2005,Cornaglia:075320:2005,BalseiroCM,Mravlje:205320:2006,Mravlje::2008}
and can be experimentally probed by conductance measurements.

We address
this subject
in the present work by investigating a
two-channel Kondo (2chK) effect
\cite{NRGtwoChKondo,Ingersent:2594:1992} in molecular systems with
``center-of-mass" vibrational modes. Two-channel Kondo physics,
originally investigated in the context of heavy-fermion materials,
\cite{Cox:1240:1987,Ingersent:2594:1992} has been an active topic
in the area of nanostructures. In semiconductor quantum dots,
several theoretical predictions \cite{QD2chKPredictions} and a
recent experimental observation\cite{Potok:169:2007} of the 2chK
have highlighted the renewed interest in such strongly correlated
states.
%
%
In addition, phonon-assisted 2chK behavior has been predicted in
effectively noninteracting systems (e.g., metallic carbon
nanotubes \cite{Clougherty:035507:2003} and
 metallic break junctions\cite{Lucignano:155418:2008}) coupled to vibronic states with
Kondo-like correlations appearing in orbital (pseudospin) degrees
of freedom. \cite{Clougherty:035507:2003,Lucignano:155418:2008}

In this paper, we consider the Kondo regime of a singly charged
molecular level connected to metallic leads, fully including both
electron-electron and electron-phonon interactions, as well as
phonon-assisted tunneling processes arising from a center-of-mass
vibronic mode. Our numerical renormalization-group (NRG)
calculations show that the presence of the extra, phonon-mediated,
conductance channel leads to non-Fermi-liquid (NFL) behavior at
low temperatures, with prominent signatures in the thermodynamic
properties and in the linear conductance. The 2chK fixed point
occurs over lines of critical points covering a wide range of
parameters, both away from particle-hole symmetry and in the
presence of deformation-induced charge-phonon couplings.

The paper is organized as follows: the model is presented in
detail in Sec.\ \ref{sec:Model} and a discussion on the 2chK
regime and the dependence of the critical parameters is given in
Sec.\ \ref{sec:2chK}. In Sec.\ \ref{sec:conductance}, we discuss
the NFL signatures in the transmission phase shift and in the
conductance across the junction. We give our concluding remarks in
Sec.\ \ref{sec:Summary}.

\begin{figure}[tbp]
\includegraphics[clip,width=0.8\columnwidth]{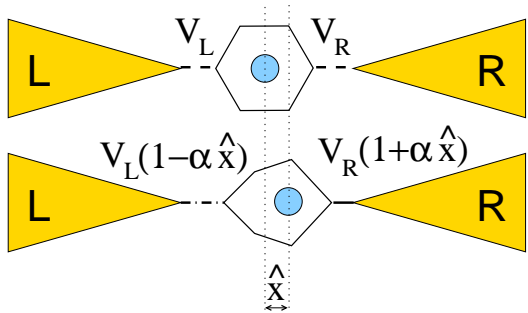}
\caption{ (color online) Schematic illustration of the
phonon-assisted tunneling process: the coupling of a molecular
level to left and right metallic leads is modulated by a
center-of-mass vibrational mode.}
\label{fig:model}
\end{figure}

\section{Model}
\label{sec:Model}

We consider a molecular complex (e.g., an organo-metallic
compound, $C_{60}$, etc.) in the Coulomb blockade (CB) regime
connected to metallic leads (for instance, in a metallic break
junction setup). Kondo correlations appear as the molecule is
tuned into a CB valley with an odd number of electrons by a
plunger gate voltage. We focus on the low-bias regime, in which
electronic transport is dominated by a singly-occupied molecular
level of energy $\epsilon_d$ (measured from the Fermi energy in
the leads and tunable by the gate voltage) with a charging energy
$U$ arising from the electron-electron interactions within the
molecule.

In our model, the molecule is connected to left (L) and right (R)
leads by tunneling couplings proportional to the overlap between
the wavefunctions of the molecular level and the (s-like) metallic
states in the leads. More importantly, we consider the effect of
phonon-mediated coupling through a center-of-mass vibrational
mode, as illustrated in Fig.\ \ref{fig:model}. For small
displacements, the molecule-lead tunneling can be effectively
written as $V_{L(R)}(1 \pm \alpha\hat{x})$ where $\hat{x}$ is a
displacement operator in the direction of the motion and $\alpha$
is a system-specific parameter, being essentially proportional to
the ratio between the overlap length and the oscillation amplitude
of the vibrational mode. Additionally, deformations in the
chemical bonding (also illustrated in Fig.\ \ref{fig:model}) will,
in general, introduce an effective coupling of the center-of-mass
oscillating mode and the charge state of the molecule. We model
this by a Holstein-type electron-phonon term with coupling
$\lambda$.

The full Hamiltonian is given by
\begin{equation}
\label{Eq:Hamiltonian}
H = H_{\rm M} + H_{\leads}  + H_{\dotleads} \; ,
\end{equation}
where
\begin{eqnarray}
\label{Eq:HamiltonianTerms}
H_{\rm M} & = & \epsilon_d n_{d \sigma} + U n_{d \uparrow}n_{d \downarrow} + \nonumber\\
& & \lambda(1-n_d)(a+a^{\dagger})+\omega_0 a^{\dagger}a \; ,\nonumber\\
H_{\dotleads} & = & \sum_{{\bf k}} V_L
\left(1-\alpha\hat{x}\right) d^{\dagger}_{\sigma} c_{L {\bf k}
\sigma}+\mbox{h.c.}
\nonumber\\
& & + V_R \left(1+\alpha\hat{x}\right) d^{\dagger}_{\sigma} c_{R
{\bf k} \sigma} + \mbox{h.c.},
\nonumber\\
H_{\leads} & = & \sum_{{\bf k},\ell=L,R} \epsilon_{\ell k}
c^{\dagger}_{\ell {\bf k} \sigma} c_{\ell {\bf k} \sigma} \; .
\end{eqnarray}

In the above, $d^{\dagger}_{\sigma}$ ($d_{\sigma}$) and
$c^{\dagger}_{\ell {\bf k} \sigma}$ ($c_{ \ell {\bf k} \sigma}$)
are fermionic operators that create (destroy) electrons with spin
$\sigma$ in the molecule and leads, respectively, ($n_{d \sigma} =
d^{\dagger}_{\sigma}d_{\sigma}$  is the electron number operator),
$\omega_0$ is the frequency of the local center-of-mass phonon
mode, with $a^{\dagger}$($a$) being the phonon operators
($\hat{x}=a+a^{\dagger}$). We assume the wide band limit and ${\bf
k}-$independent dot-lead couplings ($V_{\ell {\bf k}}\equiv
V_{\ell}$).

Hamiltonian  (\ref{Eq:Hamiltonian}) can be written as an Anderson
impurity model coupled to two independent fermionic channels.
Defining symmetric (``even") and anti-symmetric (``odd")
combinations of the electronic operators in the left and right
leads $c_{e(o) {\bf k} \sigma} \equiv \left( V_R c_{R {\bf k}
\sigma} \pm V _L c_{L {\bf k} \sigma} \right)/2\sqrt{V^2_L +
V^2_R}$, the $H_{\dotleads}$ term in (\ref{Eq:Hamiltonian})
becomes:
\beq
H_{\dotleads}  = \bar{V} \sum_{{\bf k}, \sigma}
d^{\dagger}_{\sigma} c_{e {\bf k} \sigma} +
\alpha\left(a+a^{\dagger}\right) d^{\dagger}_{\sigma} c_{o {\bf k}
\sigma}+\mbox{ h.c.} \; ,
\eeq
where $\bar{V}\equiv 2\sqrt{V^2_L + V^2_R}$.  For $\alpha\neq0$, a
phonon-mediated coupling to the odd channel is present. As we
shall see, this has important consequences in the physics of the
ground-state of the system. Notice that the odd-channel coupling
is present even for molecules not symmetrically coupled to the
leads (i.e., $V_L \neq V_R$), a more likely configuration in
experiments. For $\alpha=0$, this term vanishes and Hamiltonian
(\ref{Eq:Hamiltonian}) corresponds to the single-channel
Anderson-Holstein model, previously investigated by NRG
\cite{NRGphonons} and analytical renormalization-group methods.
\cite{Paaske:176801:2005}

The eigenstates of the Hamiltonian (\ref{Eq:Hamiltonian}) can be
labeled by total charge and $SU(2)$ spin symmetry $(Q,S)$. At the
molecule site, states are also labeled by the number of phonons
$m$ $\left(\left(a^{\dagger}\right)^m|0\rangle=|m\rangle \right)$
Notice that both the electron-phonon  ($\propto \lambda$) and the
phonon-assisted tunneling ($\propto \alpha$) terms couple
$|m\rangle$ and $|m\pm1\rangle$ states. This last term couples
only states with a difference of one electron in the odd channel
and, thus, commutes with a generalized \textit{parity} operator,
defined by $\hat{P}=(-1)^{m+Q_o+1}$ where $Q_o$ is the total
charge in the odd channel. Therefore, for $\lambda=0$, the
Hamiltonian (\ref{Eq:Hamiltonian}) has an additional $O(1)$ parity
symmetry (which is lost for $\lambda \neq 0$, as the
electron-phonon term will couple states with different parity).
\begin{figure}[tbp]
\includegraphics[clip,width=1.0\columnwidth]{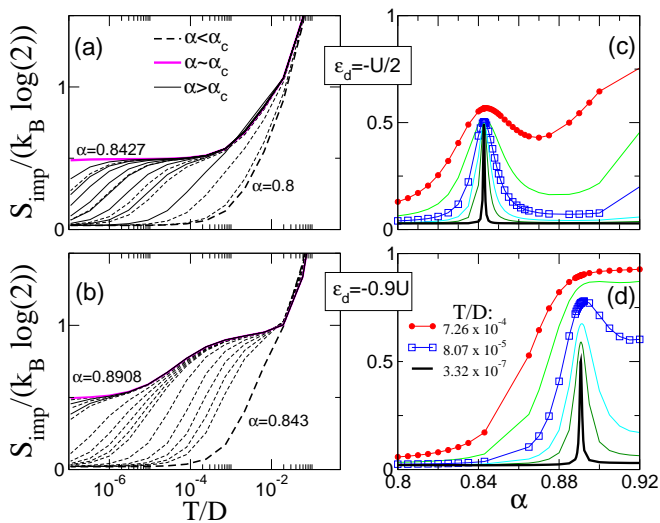}
\caption{ (color online) Impurity contribution to the entropy
$S_{\imp}$ at the particle-hole symmetric
point $\epsilon_d=-U/2$~(a,c) and at $\varepsilon_d=-0.9U$~(b,d). In both cases,  an
unstable 2chK fixed point for which $S_{\imp} \rightarrow k_B
\log{2^{1/2}}$  as $T \rightarrow 0$ is reached at
$\alpha=\alpha_c$ . Panels (a) and (b) show $S_{\imp}$ versus
$T/D$ for fixed $\alpha$ while (c) and (d) show $S_{\imp}$ versus
$\alpha$ for fixed temperatures. }
\label{fig:thermo}
\end{figure}

We solve the two-channel problem with Wilson's numerical
renormalization-group  (NRG) technique\cite{Bulla:395:2008}
adapted to include the phonon degrees  of freedom.
\cite{NRGphonons} As it is standard in the NRG method, we map Eq.\
(\ref{Eq:Hamiltonian}) into a (two-band) tight-binding Hamiltonian
by performing logarithmic discretizations of the continuum spectra
in the even and odd fermionic channels. In the calculations, we
use a discretization parameter $\Lambda=3$ and kept up to 2500
states in the NRG iterations, which proves to be adequate for the
calculation of thermodynamic properties. We have used a cutoff of
$N_{\rm ph}=9$ in the maximum number of phonons and checked for
convergence of the results
with $N_{\rm ph}$. We assume a constant (metallic) density of
states $\rho_0$ in the leads with bandwidth $D$.

\section{Two-channel Kondo physics}
\label{sec:2chK}

From the NRG spectra, we calculated thermodynamic properties for
this system. A particularly revealing quantity is the contribution
to the total entropy coming from the ``impurity" degrees of
freedom (in the present case, the molecule), defined as the
difference between the total entropy $S(T)$ and the entropy
calculated in the absence of the molecule $S^{(0)}(T)$. As it has
been shown by Bethe-ansatz\cite{Bolech:237206:2002} and NRG
calculations\cite{Anders:121101:2005} in two-channel Kondo models,
$S_{\imp}(T)\equiv S(T)-S^{(0)}(T)$ reaches a universal
low-temperature plateau at $S_{\imp}/k_B=1/2 \ln(2)$ ($k_B$ is
Boltzmann's constant) when both channels are equally coupled to
the impurity.

 Fig. \ref{fig:thermo} shows  $S_{\imp}$ versus temperature
$T/D$ and $\alpha$ for two values of the molecular level energy:
$\varepsilon_d=-U/2$, corresponding to the particle-hole symmetric
point (top panels), and $\varepsilon_d=-0.9U$ (bottom panels).
\cite{ParamValues} In both cases, an entropy plateau
$S_{\imp}/k_B=1/2 \ln(2)$ is reached at low temperatures (Figs.\
\ref{fig:thermo}(a) and (b)), signaling the presence of a
non-Fermi liquid (NFL) fixed point.

This NFL fixed point is reached as $\alpha$ approaches a critical
value $\alpha_c$ and at temperatures below a characteristic
crossover energy scale $T^{*}$.  Deviations from $\alpha=\alpha_c$
drive the system away from this state to the more conventional
Kondo-screened state (characterized by $S_{\imp}=0$), illustrating
the unstable nature of the fixed point. This is depicted in Fig.\
\ref{fig:thermo}(c), showing $S_{\imp}$ versus $\alpha$ at fixed
temperatures. At lower temperatures, a narrow peak   of height
$S_{\imp}=1/2 \ln{2}$ pinpoints  the critical value
$\alpha=\alpha_c$. This marks the position of the fixed point in
parameter space. At higher temperatures, the broadening of these
peaks indicate that signatures of NFL behavior extend over a
finite range of $\alpha$.

This behavior persists away from particle-hole symmetry, as shown
in Figs.\ \ref{fig:thermo}(b) and (d). Interestingly, the critical
value $\alpha_c$ \textit{increases} as the system approaches the
mixed-valence regime ($\varepsilon_d \rightarrow 0,-U$; $\langle
n_d \rangle \rightarrow 0,2$) as compared with the $\varepsilon_d
=-U/2$ case ($\langle n_d \rangle=1$). This contrasts with the
(phonon-independent) two-channel Anderson model (2chAM), for which
the critical couplings are $\varepsilon_d$-independent.
\cite{Anders:121101:2005} The crossover energy scale to the NFL
fixed point, related to the Kondo temperature, decreases
exponentially as the system enters the mixed-valence regime, as in
the 2chAM.

In the mixed-valence regime, the system flows into a Fermi-liquid
fixed point with $S_{\imp}=k_B \ln(2)$ at higher temperatures
before entering the NFL regime (as seen, e.g., in
Fig.\ref{fig:thermo}(b)). This fixed point is nonmagnetic and
characterized by a \textit{parity} degeneracy in the ground state
(rather than the usual spin degeneracy in similar models).
\begin{figure}[tbp]
\includegraphics[clip,width=1.0\columnwidth]{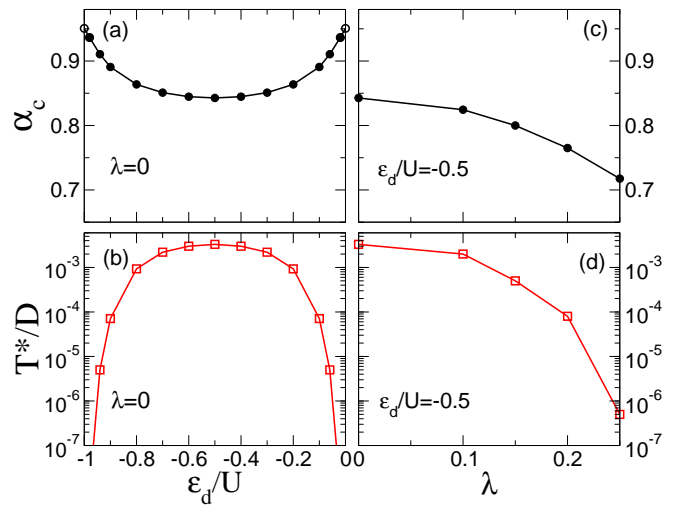}
\caption{ (color online) Critical dot-lead coupling $\alpha_c$ (a)
and crossover temperature $T^{*}$ (b) vs. molecule level position
$\varepsilon_d$. (c,d) Same quantities but now as a function of the
electron-phonon coupling $\lambda$ at $\varepsilon_d=-U/2$. }
\label{fig:alphac}
\end{figure}

As previously discussed, the $S_{\imp}$ vs.  $T/D$ curves display
the NFL plateau at low temperatures for $-U < \varepsilon_d < 0$
at a critical $\alpha=\alpha_c$. The critical value  $\alpha_c$
increases as the system is moved away from particle-hole symmetry,
reaching its highest values at $\varepsilon_d = 0^-,-U^+$, as
depicted in Fig.\ \ref{fig:alphac}(a). In addition, at
$\alpha=\alpha_c$ the crossover temperature sharply decreases as
the systems is driven away from the particle-hole symmetric point.
This is illustrated in Fig.\ \ref{fig:alphac}(b), where we define
the crossover temperature $T^*$ as $S_{\imp}(T=T^*)=3/4 \ln{2}$.
This panel shows that $T^*$ decreases exponentially as
$\varepsilon_d \rightarrow -U,0$, indicating that the 2chK is only
reached within the range of gate voltages for which the molecular
level in singly occupied. In fact, we find the NFL fixed point in
this ``local moment" range ($-U < \varepsilon_d < 0$) only,
indicating, along with susceptibility calculations, that the Kondo
screening occurs in the ``real spin" as opposed to a ``pseudospin"
degree of freedom (e.g., those connected to charge states or
$|m\rangle$,$|m \pm1\rangle$ phonon states). In this range, the
value $\alpha_c$ varies nearly quadratic with $\varepsilon_d$, as
shown in Fig.\ \ref{fig:alphac}(a).
\begin{figure}[tbp]
\includegraphics[clip,width=1.0\columnwidth]{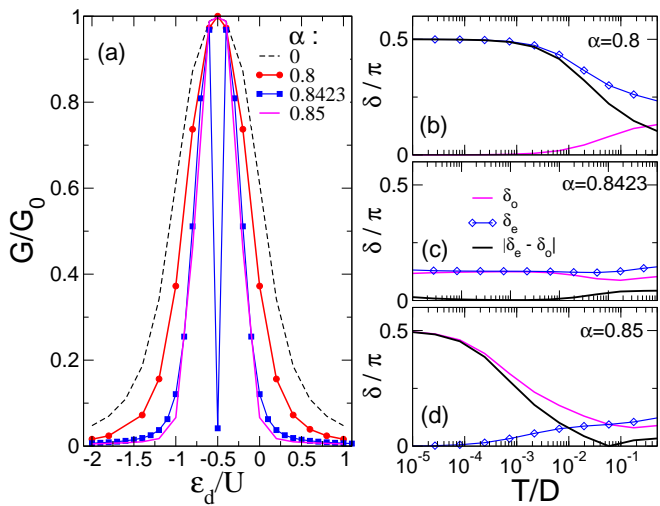}
\caption{ (color online) (a) Conductance $G$ versus
$\varepsilon_d$ for different values of $\alpha$. A dip appears
for $\alpha=\alpha_c$ at $\varepsilon_d=-U/2$, signaling the 2chK
state. (b-d) Even ($\delta_e$), odd ($\delta_o$) and relative
$|\delta_e-\delta_o|$ phase shifts vs. temperature at
$\varepsilon_d=-U/2$ and different values of $\alpha$.  }
\label{fig:conductance}
\end{figure}

Furthermore, for fixed $\varepsilon_d$, both $\alpha_c$ and
$T^{*}$ \textit{decrease} as the electron-phonon coupling
$\lambda$ increases (Figs.\ \ref{fig:alphac}-(c) and (d)). The
decrease in $T^{*}$ is consistent with the fact that the
electron-phonon coupling renormalizes  the electron-electron
interaction as $U_{\mbox{\scriptsize eff}} \approx
U-2\lambda^2/\omega_0$, \cite{NRGphonons} effectively driving the
system into the mixed valence regime. The decrease in $\alpha_c$
with $\lambda$ indicates that not only the molecule-lead couplings
but also the coupling of the vibrational mode with the
\textit{charge state} of the molecule plays a role in the
mechanism leading to the two-channel Kondo effect.

\section{Phase shifts and conductance}
\label{sec:conductance}

We now turn to the transport properties across the molecule. We
first point out that  the presence of the CM phonon term in Eq.\
(\ref{Eq:Hamiltonian}) breaks the ``proportionate coupling''
condition\cite{Meir:2512:1992} between left and right leads (even
in a rather unrealistic symmetric coupling configuration
$V_L=V_R$). For this reason, a calculation of the linear
conductance via a Landauer-like formula\cite{Meir:2512:1992} would
involve not only the local interacting retarded Green's function
(obtainable from NRG) but also Keldysh Green's functions.

Instead, we turn  to the equivalent approach of calculating the
$T=0$ conductance using the scattering phase shifts,
\cite{Pustilnik:216601:2001,Lucignano:155418:2008,Mravlje::2008}
as $G=G_0 \sin^2{\left(\delta_e - \delta_o\right)}$, where
$\delta_e$ and $\delta_o$ are the phase shifts in the even and odd
channels, respectively, and
$G_0=(2e^2/h) \sin^2{\theta}$ with $\theta$ being an overall phase
that depends on the microscopic details of the molecule-lead
junction (we henceforth consider $\theta=\pi/2$).

The phase shifts $\delta_{e(o)}$ can  be obtained from the  NRG
spectra \cite{Affleck:7918:1992,Hofstetter:235301:2004} using the
$(Q,S,P)$ quantum numbers to label the states.
\cite{Lucignano:155418:2008}  We note that the parity  quantum
number is strictly  conserved only for $\lambda=0$. Although it is
possible,      in    some   cases,   to      calculate  the
difference $\delta_{e}-\delta_{o}$ from the $(Q,S)$ NRG
spectra,\cite{Hofstetter:235301:2004,Mravlje::2008} in the
following we use  $\lambda=0$ as it  retains most  of the
interesting physics . We should note that, away from FL fixed
points, the correspondence between the excitations in the NRG
spectra and the phase shifts entering the conductance formula is
only approximate. Nevertheless, we expect the conductance obtained
with this prescription to give a qualitatively accurate picture in
the NFL state as well, as discussed below.

Results    for    the    conductance    are    shown    in Fig.\
\ref{fig:conductance}(a).   For   $\alpha=0$, the  familiar shape
is recovered: $G=G_0$   at  the    particle-hole symmetric point
$\varepsilon_d=-U/2$ an $G \rightarrow 0$ as $|\varepsilon_d|$
increases. As $\alpha$ increases, the peak narrows with $G=G_0$ at
the p-h symmetric point for $\alpha<\alpha_c$ and
$\alpha>\alpha_c$. Interestingly, as $\alpha$ approaches  the
critical value $\alpha_c$ at $\varepsilon_d=-U/2$,  a \textit{dip}
appears in the conductance curve.

This is a indication of the NFL behavior and a signature of the
two-channel   fixed    point.  The behavior of individual phase
shifts at the particle-hole symmetric point the is illustrated in
Figs.\ref{fig:conductance}(b,c,d). For $\alpha < \alpha_c$
(Fig.\ref{fig:conductance}(b)), $\delta_e \rightarrow \pi/2$ and
$\delta_o \rightarrow 0$ at low temperatures, indicating a
decoupling of the odd channel, while for $\alpha > \alpha_c$ the
odd channel becomes strongly coupled at low temperatures
($\delta_o \rightarrow \pi/2$, $\delta_e \rightarrow 0$ in
Figs.\ref{fig:conductance}-(d)). In both cases, one expects a peak
in the conductance. \cite{Pustilnik:216601:2001}

At the critical point $\alpha=\alpha_c$ (NFL regime), the NRG
spectra is identical for both even and odd parities, as predicted
by conformal field theory. \cite{Affleck:7918:1992} In this case,
our prescription for obtaining the phase shifts gives
$\delta_e=\delta_o$ at low temperatures, as depicted in Fig.\
\ref{fig:conductance}(b), causing a destructive interference and
suppressing the transmission. We thus expect this result to hold,
even though the individual values of $\delta_{e(o)}$ obtained from
the NRG spectra in the NFL regime are only approximate.

These signatures in the low temperature conductance versus gate
voltage curves can, in principle, identify the 2chK regime in
molecular junctions. Experimentally, a fine tuning of the
microscopic parameter $\alpha$ to the critical value is
nonetheless a challenging task. In general, the value of $\alpha$
will be determined by specific details of the junction, such as
the ratio of amplitude of the center-of-mass vibration  and the
length of the molecule-lead overlap. It is interesting, however,
to notice that, at higher temperatures, one might obtain $\delta_e
\approx \delta_o$ over a wider range of $\alpha$ and
$\varepsilon_d$ near the critical values, as illustrated in Figs.\
\ref{fig:thermo}(c)-(d). In this case, one expects a
finite-temperature signature of the $T=0$ non-Fermi liquid point.

\section{Summary}
\label{sec:Summary}

In summary, we have studied center-of-mass vibrational effects and
phonon-assisted processes in the transport properties of a
molecular junction in the Kondo regime. The interplay between
electron-electron and electron-phonon interactions in this system
can be described by an effective two-channel Anderson model with
phonon-assisted tunnel couplings. Our numerical
renormalization-group calculations for the thermodynamic
properties of the effective model show non-Fermi-liquid effects
below a characteristic crossover temperature over critical lines
in parameter space.

We find that the crossover temperature is at a maximum at the
particle-hole symmetric point and rapidly approaches zero as the
system enters the mixed valence regime. Furthermore, we find
distinct signatures of the non-Fermi-liquid phase in the linear
conductance.

\acknowledgments

We thank K. Al-Hassanieh, G. Martins, C. B\"usser, K. Ingersent,
A. Feiguin, and M. Daghofer for fruitful discussions.
We acknowledge support from NSF grant DMR-0706020. Research at
ORNL is sponsored by the Division of Materials Sciences and
Engineering, Office of Basic Energy Sciences, U.S. Department of
Energy, under contract DE-AC05-00OR22725 with Oak Ridge National
Laboratory, managed and operated by UT-Battelle, LLC.


\begin{thebibliography}{10}
\bibitem{ParamValues}
For concreteness, we set $\omega_0=0.2D$, $U=0.5D$ and
$\bar{V}=0.2D$ although qualitatively similar results were
obtained using different values of these parameters.
\end{thebibliography}
\end{document}